\documentclass[sn-aps]{sn-jnl}% American Physical Society (APS) Reference Style
\usepackage{graphicx}%
\usepackage{multirow}%
\usepackage{amsmath,amssymb,amsfonts}%
\usepackage{amsthm}%
\usepackage{mathrsfs}%
\usepackage[title]{appendix}%
\usepackage[table]{xcolor}%
\usepackage{textcomp}%
\usepackage{manyfoot}%
\usepackage{booktabs}%
\usepackage{algorithm}%
\usepackage{algorithmicx}%
\usepackage{algpseudocode}%
\usepackage{listings}%
\usepackage{longtable}%
\usepackage{lscape}
\begin{document}

\title[]{Shifting the Paradigm: Estimating Heterogeneous Treatment Effects in the Development of Walkable Cities Design}

\author[1]{\fnm{Jie} \sur{Zhu}}\email{jie.zhu@sydney.edu.au}
\equalcont{These authors contributed equally to this work.}

\author*[2]{\fnm{Bojing} \sur{Liao}}\email{lbj@xmu.edu.cn}
\equalcont{These authors contributed equally to this work.}

\affil[1]{\orgdiv{School of Business}, \orgname{University of Sydney}, \orgaddress{\city{Sydney}, \postcode{2000}, \state{NSW}, \country{Australia}}}
\affil[1]{\orgname{Willow Inc}, \orgaddress{\city{Sydney}, \postcode{2000}, \state{NSW}, \country{Australia}}}

\affil*[2]{\orgdiv{Institute of Creativity and Innovation}, \orgname{Xiamen University}, \orgaddress{\street{No.422 Siming South Road}, \city{Xiamen}, \postcode{361000}, \state{Fujian}, \country{China}}}


\abstract{
The transformation of urban environments to accommodate growing populations has profoundly impacted public health and well-being. This paper addresses the critical challenge of estimating the impact of urban design interventions on diverse populations. Traditional approaches, reliant on questionnaires and stated preference techniques, are limited by recall bias and capturing the complex dynamics between environmental attributes and individual characteristics. To address these challenges, we integrate Virtual Reality (VR) with observational causal inference methods to estimate heterogeneous treatment effects, specifically employing Targeted Maximum Likelihood Estimation (TMLE) for its robustness against model misspecification. Our innovative approach leverages VR-based experiment to collect data that reflects perceptual and experiential factors. The result shows the heterogeneous impacts of urban design elements on public health and underscore the necessity for personalized urban design interventions. This study not only extends the application of TMLE to built environment research but also informs public health policy by illuminating the nuanced effects of urban design on mental well-being and advocating for tailored strategies that foster equitable, health-promoting urban spaces.
}




\maketitle

\section{Introduction}\label{sec1}

Urbanization has dramatically transformed environments and lifestyles, becoming a focal point for accommodating a growing population\cite{collins2024making,loo2017neighborhood}. It has been an intriguing topic for scholars to understand the intricate relationship between individual well-being and the socio-cultural environment since the 20th century \cite{collins2024making}. Recent focus on urban design's role in addressing global public health issues marks a resurgence of interest in this field\cite{loo2017neighborhood,giles2016city}. Research underscores impact of the built environment on health, mental wellness, and life quality, noting that neighborhood designs promoting physical activity, green space, and aesthetics are crucial for combating obesity, heart disease, and mental health issues\cite{nieuwenhuijsen2022evaluation,doiron2020healthy,hajna2018laboratory,houlden2019spatial, crouse2019complex}. Specifically, the benefits of walkable neighborhood for mental health and social well-being are well-documented, underscoring the importance of creating environments that support health and sustainable communities\cite{kellert1993biophilia,felsten2009take,giles2016city,sallis2016physical}.

However, estimating the potential effects of urban (built) environment interventions on diverse populations remains a complex challenge. Traditionally, researchers have utilized a variety of methodological approaches to explain the causal impact of the built environment on mental health outcomes. This broad spectrum of methodologies reflects the nuanced and complex nature of how environmental contexts influence subjective well-being. Among these approaches, questionnaires and surveys have been extensively employed to evaluate individuals' perceptions, experiences, and self-reported mental health status, particularly in relation to the urban environments within their neighborhood. Such data is invaluable for assessing how personal interpretations of environmental factors correlate with mental health outcomes \cite{leslie2010perceived,mytton2012green}. 

Although these methods are instrumental in gathering large sets of subjective data, they are not without limitations. A significant concern associated with questionnaire and surveys is their susceptibility to recall bias \cite{liao2022individuals}. Recall bias occurs because traditional tools depend heavily on the respondents' ability to accurately remember and report past events or experiences. This type of bias can particularly skew the data when individuals may not accurately remember or may alter their recollections--intentionally or unintentionally --based on current feelings or misconceptions \cite{desjardins2023improving}. For example, a recent study by Li et al. (2022) highlights how recall bias can affect research findings, particularly in studies that rely on self-reported stressful incidents or overemphasize certain events depending on their current mood or recent experiences, thus distorting the true impact of the built environment on their psychological state \cite{li2022environmental}.
 
To overcome the limitations inherent in transitional survey methods, the conjoint experiments and the stated preference method have been employed to elicit individuals' preferences and hypothetical choices concerning urban built environment scenarios with varying levels of attributes \cite{kaczynski2007environmental}. By presenting respondents with hypothetical scenarios featuring different combinations of environmental characteristics, these methods can provide insights into the relative importance of various factors in decision-making processes concerning urban environments. For example, in a study by Zhao et al. (2022), participants were presented with a series of hypothetical residential scenarios differing in proximity to green spaces and types of amenities available \cite{zhao2022comparing}. The researchers used a conjoint analysis approach to quantify how each attribute influenced the participants' preferences for one living scenario over another, revealing a strong preference for proximity to green spaces over other factors. 

Nevertheless, while conjoint experiments and stated preference methods offer valuable insights, they also have notable limitations. One such limitation is their inability to fully account for the intricate interplay between environmental factors, such as the presence of green spaces and land use mix-diversity, and individual characteristics like socioeconomic status and physical health. Furthermore, these methods often do not capture actual behavioral responses in real-world settings, potentially leading to discrepancies between stated and actual preferences \cite{zhao2022comparing}. For instance, Hurtubia et al. (2021) explored the preferences for bike sharing stations in the perceptions of public spaces using stated preference methods \cite{hurtubia2021role}. While the results indicated a positive preference for presence of bike sharing, subsequent observational studies in the same urban areas showed a negative preference for disorganized dock-less bikes on sidewalks. It means that actual behavior might diverge from hypothetical choices due to factors not captured in the initial surveys, such as actual accessibility, perceived safety, and personal time constraints \cite{liao2022individuals,birenboim2019utilization}. 

To address these challenges, we propose a novel approach that integrates Virtual Reality (VR) technology with observational causal inference methods to estimate heterogeneous treatment effects of urban built environment interventions on mental health outcomes. The primary objective of causal inference is to quantify the impact of a specific intervention. This concept is grounded in the counterfactual framework initiated by Rubin \cite{rubin2005causal}. A notable development in this field is the Targeted Maximum Likelihood Estimation (TMLE) \cite{van2006targeted}, introduced by Van der Laan and Rubin in 2006. TMLE stands out from traditional methods by utilizing both the treatment-generating mechanism (propensity score model) and the outcome-generating mechanism to estimate the average treatment effect, making it doubly robust. This means it remains reliable even if either the exposure or outcome model is incorrectly specified, and it has been proven more effective than other methods like inverse probability of treatment weighting and propensity score matching, especially in cases of likely model misspecification.

Despite its growing recognition, the application of TMLE for Likert outcomes is scarce. Likert outcomes are common in sociology but are often simplified to latent groups for ease of analysis, which can lead to loss of information and variable statistical power. Understanding TMLE in the context of Likert outcomes is crucial due to these complexities. Most existing TMLE guides focus on simulated data, but this article aims to demonstrate TMLE using a real-world dataset, helping practitioners understand its application in practical scenarios and highlighting its differences from other commonly used methods. To bridge the gap between theory and practice, we present a case study that illustrates the nuances of applying TMLE to a real-world situation. In doing so, we provide a VR-based conjoint experiment example of how TMLE can be used to interpret Likert data within the rich, multifaceted domain of social research.

The VR-based conjoint experiment has revolutionized the way researchers investigate individual preferences and behavioral responses regarding urban environments \cite{liao2022individuals,zhao2022comparing,hayek2016bringing}. This advanced methodology utilizes realistic three-dimensional simulations of neighborhoods, which engage participants in a highly immersive and interactive manner. By doing so, VR-based conjoint experiments provide a controlled and replicable platform for data collection, which is particularly adept at capturing the nuanced influence of perceptual and experiential factors on human behavior \cite{kasraian2021evaluating}. Traditional survey methods often fail to capture the full sensory and emotional responses that individuals might have to real-world environments \cite{kasraian2021evaluating, liao2022individuals, birenboim2019utilization, zhao2022comparing}. VR overcomes this limitation by providing a rich, multi-sensory experience that can include visual and auditory, thereby mimicking real-life experiences more closely \cite{birenboim2019utilization}. This allows researchers to observe how these complex perceptual inputs influence decision-making and preference formation in a way that abstract surveys cannot \cite{kasraian2021evaluating, birenboim2019utilization}.

The VR-based conjoint experimental design is particularly suited to the application of TMLE, as it generates the type of complex, multi-layered data for which TMLE was developed to analyze. By leveraging the TMLE framework with a joint treatment approach (e.g., accounting for correlations between environmental factors), our methodology enables estimating individualized treatment effects while accounting for the complex interplay between multiple environmental factors and individual characteristics\cite{schuler2017targeted}. This approach not only enhances the accuracy of the resulting data analysis but also offers a more nuanced understanding of the effects of urban design interventions.

This study reveals meaningful insights, establishing that interventions such as land use mix (LM), open spaces (OS), block connectivity (BC), Road Size (RS), and green/trees (GT) exert a positive effect on Conditional Average Treatment Effect (CATE) values, with OS interventions exhibiting notable efficacy as standalone measures. The research highlights the complexity inherent to urban walkability and calls for a judicious interpretation of CATE values in light of possible confounding due to demographic variables. The evidence points to the conclusion that deliberate urban design improvements, especially those integrating green and open spaces, are instrumental in enhancing walkability. While LM and BC interventions lead to moderate gains, the impact of Road Size (RS) interventions invites reconsideration, as they may detract from walkability if applied without complementary measures. These pivotal findings advocate for a strategic urban design methodology that leverages intervention benefits to foster environments conducive to pedestrians.

Our study contributes significantly to both theoretical advancements and practical applications in this field. First, we expand the application of TMLE and joint treatment estimation to the innovative context of urban built environment interventions. This advancement enhances the field of causal inference, deepening our understanding of the nuanced relationships between the urban built environment and mental health outcomes.
Second, our findings provide invaluable insights for urban design and public health policies. By facilitating the development of tailored interventions that cater to diverse population segments, our study optimizes resource allocation and maximizes positive impacts on mental well-being. 
Finally, our work lays the groundwork for creating equitable and sustainable urban built environments that promote holistic well-being for all. It challenges one-size-fits-all approaches and underscores the importance of personalized strategies.

\section{Methods}\label{sec11}

\subsection{Case Study Design} 

In our study, we employed a virtual reality (VR)-based conjoint experiment to assess individuals' preferences and behaviors in response to urban design interventions. This approach features three stages: specifying attributes and their levels, designing 3D neighborhood simulations, and developing an online questionnaire to capture perceptual and experiential responses within a VR environment. 

The choice of attributes and levels of VR-based conjoint experiment design were based on earlier works\cite{liao2022individuals}. The experiment focuses on the street block level, chosen for its analogous influence on walking behavior to that of a neighborhood, as supported by literature\cite{sallis2009measuring}. This scale allows for detailed 3D modeling and enables participants to closely engage with the urban environment's features. 

Attributes impacting walking behavior—land use mix-diversity, walking facilities, sidewalks, and trees—were identified based on prior research, along with connectivity and open space\cite{liao2022individuals, liao2020empirical}. Five attributes were selected for the street block experiment, each with two levels: land use mix (exclusively residential or mixed-use), block connectivity (high or low), road size (two lanes with narrow pedestrian zones or one lane with wide pedestrian zones), open space (presence or absence), and green/trees (presence or absence of trees).

From the potential 32 ($2^{5}$) attribute combinations, we applied a fractional factorial design to reduce the number while preserving orthogonality, ensuring statistical independence among attributes. This approach resulted in eight distinct attribute profiles, facilitating efficient estimation of main effects without attribute correlation\cite{hensher2005applied}. This design strategy enhances our experimental setup's capacity to discern the individual contributions of various urban design elements to walking behavior, streamlining the process of identifying effective interventions for promoting walkability.

\begin{table}[!h]
\caption{An orthogonal design of eight attribute profiles.}
\label{table_2}
\begin{tabular}{lllllllll}
\textbf{Attributes} & \textbf{Set 1}            & \textbf{Set 2}            & \textbf{Set 3}            & \textbf{Set 4}            & \textbf{Set 5}            & \textbf{Set 6}            & \textbf{Set 7}            & \textbf{Set 8}            \\
\hline
Land use mix        & \cellcolor{gray!10}1 & \cellcolor{gray!10}1 & \cellcolor{gray!10}1 & \cellcolor{gray!10}1 & 0                         & 0                         & 0                         & 0                         \\
Block connectivity  & \cellcolor{gray!10}1 & \cellcolor{gray!10}1 & 0                         & 0                         & \cellcolor{gray!10}1 & \cellcolor{gray!10}1 & 0                         & 0                         \\
Road size           & \cellcolor{gray!10}1 & \cellcolor{gray!10}1 & 0                         & 0                         & 0                         & 0                         & \cellcolor{gray!10}1 & \cellcolor{gray!10}1 \\
Open space          & \cellcolor{gray!10}1 & 0                         & \cellcolor{gray!10}1 & 0                         & \cellcolor{gray!10}1 & 0                         & \cellcolor{gray!10}1 & 0                         \\
Green/Trees               & \cellcolor{gray!10}1 & 0                         & 0                         & \cellcolor{gray!10}1 & 0                         & \cellcolor{gray!10}1 & \cellcolor{gray!10}1 & 0          \\
\hline
\end{tabular}
\end{table}

In the subsequent phase, the eight attribute combinations were translated into corresponding virtual reality (VR) environments. To accomplish this, we constructed the foundational 3D model to represent the Dutch street block by using SketchUp Pro. The dimensions of the experiment area within the street block spanned 300 meters in length and 240 meters in width. While maintaining a consistent road width, we introduced variations in road type: (1) two lanes dedicated to cars with a narrow pedestrian sidewalk, and (2) a single lane for cars with a wider pedestrian pathway.

Regarding the land use mix attribute, we established two levels: the first level entailed an exclusively residential street block, while the second level incorporated a mix of residential and commercial areas such as shops and supermarkets dispersed within the residential area. In terms of the block connectivity attribute, we manipulated the number of intersection points within the street block. Additionally, we introduced variations in the presence or absence of open space, as well as the inclusion or exclusion of street trees to represent the green attribute. By utilizing the 3D sketch models and altering attribute levels accordingly, we generated a total of eight distinct VR environments.

Subsequently, all eight VR environments were imported into Twinmotion, a rapid 3D rendering software\cite{wu2020design}. Within Twinmotion, we integrated materials, trees, traffic elements, facilities, and human figures into each of the 3D sketch models, enhancing their realism and creating more interactive experiences. Then, we established a consistent walking perspective and exported all virtual reality environments as video files. To ensure coherence, every video depicting the virtual reality environments adhered to identical parameters, including the walking route, viewing direction, geographical location, duration of sunlight, seasonal conditions, and weather conditions. Each video had a standardized duration of 90 seconds.

Furthermore, to maintain uniformity in the conveyed information, all scenarios incorporated identical 3D objects, such as buildings, trees, and facilities, to represent the respective characteristics. Consequently, participants were presented with consistent attributes throughout all scenarios, including uniform tree colors, building styles, and facility materials. This approach ensured that participants evaluated the environments based on the attributes utilized in their construction, while keeping all other factors constant.

The questionnaire employed in this study is divided into two distinct sections. The first part focuses on individuals' perceptions of their existing neighborhood and personal characteristics, while the second part encompasses the virtual reality (VR) environment, comprising videos and related questions intended to elicit participants' perceptions of the VR environments. Within the VR environments section of the questionnaire, participants are prompted to evaluate their experiences of the virtual environments while viewing the corresponding videos. They are asked to provide ratings based on the emotions evoked by each video. To manage the length of the questionnaire, four out of the eight dynamic 3D videos showcasing the VR environments are randomly presented to each respondent. The participant's perception of each virtual reality environment is assessed through two sections of questions.

The first section encompasses two inquiries pertaining to the environment's quality, specifically: (1) "How satisfied are you with the overall quality of this virtual environment?"; and (2) "How satisfied are you with the walking friendliness of this virtual environment?" Participants are asked to rate their satisfaction using a 7-point Likert scale, ranging from "not at all satisfied" to "fully satisfied." To capture the participants' feelings during the virtual walk-through experience, the preference rating method introduced by Birenboim et al.(2019) is employed\cite{birenboim2019utilization}. The second section of the questionnaire focuses on the emotional responses evoked by the virtual environment. Four dimensions of emotions associated with perceived walkability during the virtual walk-through experience are examined: happiness, comfort, annoyance, and security. Participants are requested to indicate the extent to which they experienced each of these emotions. The questions are presented as statements, such as "I felt happy/comfortable/annoyed/secure." For each item, respondents provide their responses on a 7-point Likert scale, ranging from "completely disagree" (1) to "completely agree" (7), as illustrated in Table 5. Additionally, the second section includes inquiries concerning the perceived benefits derived from the virtual environment.

The data for this study was sourced from earlier works by Liao et al. (2022)\cite{liao2022individuals}. Participants were recruited for this study from a nationally representative consumer panel in the Netherlands, as well as through popular social media platforms such as Twitter (\textit{X}) , LinkedIn, and Facebook. Upon introducing the virtual reality (VR) environments, respondents were informed that they would be presented with scenarios depicting a typical Dutch street block in a virtual setting. Subsequently, participants were requested to evaluate the overall quality and walking friendliness of the virtual scenarios, as well as provide their subjective emotional responses while viewing these scenarios.

A total of 308 individuals completed the online questionnaire, with 272 respondents sourced from the consumer panel and an additional 36 participants recruited through social media channels. To ensure the reliability and robustness of the data, respondents who provided repetitive answers to each question or who completed the VR portion of the questionnaire in less than 8 minutes were excluded from the analysis. Following the data cleaning process, a final sample of 295 respondents remained for further analysis. All participants were exposed to four 3D-videos, resulting in a total of 1,180 ratings recorded for each item of interest. Consequently, the final dataset consists of observations for each respondent pertaining to 4 virtual walking trips.

\subsection{Conditional Average Treatment Effect Estimcation} 
Rubin's potential outcomes framework \cite{rubin2005causal} suggests that in studies with a binary treatment, we can only observe the actual outcomes for each subject under the specific treatment they received. To understand a treatment's effect, we must compare what actually happened to a subject with the treatment to what would have happened without it or with a different treatment. This hypothetical scenario is called the \textit{counterfactual outcome}. The challenge, especially in observational studies, is accurately estimating these counterfactual outcomes. This difficulty arises because the characteristics of the groups receiving different treatments can vary significantly, leading to biased comparisons. In simpler terms, certain traits may make some people more likely to receive a specific treatment, complicating the task of isolating the treatment's true effect.

We formalize our problem using a sample dataset which has $n$ independent and identically distributed examples $ (X_i, Y_i, W^{\eta}_i) $, $ i = 1, \ldots, n $, where $ X_i \in \mathcal{X} $ represents individual features, $ Y_i \in \mathbb{R} $ is the observed outcome, and $ W^{\eta}_i \in \{0, 1\} \text{ and } \eta = 1,2, \ldots, H$ stands for $H$ treatment options and their assignment status. We posit the existence of potential outcomes $ \{Y_i(W^{\eta}_i=1), Y_i(W^{\eta}_i=0)\} $ corresponding to the outcome we would have observed given the treatment assignment $ W^{\eta}_i=1 $ or $ W^{\eta}_i=0 $ respectively, such that $ Y_i = Y_i(W^{\eta}_i) $.
One common quantity researchers interested is the average treatment effect for treatment $\eta$, which is $ \tau = \mathbb{E} [Y_i(W^{\eta}=1) - Y_i(W^{\eta}=0)] $. Here, in contrast, we want to understand how treatment effects vary with the observed covariates $ X_i $, and consider the conditional average treatment effect (CATE):
\begin{equation}
    \psi(X)^{\eta} = E_{X=x}\bigg(E(Y|W^{\eta}=1, X=x) - E(Y|W^{\eta}=0, X=x)\bigg).
\label{equation1}
\end{equation}

\noindent When the effect of multiple joint treatments are interested, we can extend the definition of CATE : 
\begin{align}
\psi(X)^{\eta_1,\eta_2, \ldots,\eta_H} = E_{X=x}\bigg(E(Y|W^{\eta_1}=1,W^{\eta_2}=1, \ldots,W^{\eta_H}=1,X=x) -\\ E(Y|W^{\eta_1}=0,W^{\eta_2}=0, \ldots,W^{\eta_H}=0,X=x)\bigg)
\end{align}

\noindent When the effect of a subset of joint treatments are interested, we can define CATE as:
    
\begin{align*}
\psi(X)^{\eta_m,\eta_n} = E_{X=x, \eta_k=\mathbb{H}}\bigg(E(Y|W^{\eta_m}=1,W^{\eta_n}=1, W^{\eta_k}=\mathbb{H},X=x)-\\ E(Y|W^{\eta_m}=0,W^{\eta_n}=0, W^{\eta_k}=\mathbb{H},X=x)\bigg),
\end{align*}
\noindent where $m, n \in \{1,2, \ldots, H\}$ are the treatment options of interest, $\eta_k=\mathbb{H}, k \in \{1,2, \ldots, H\}$ and $k \notin \{m,n\} $ stands for observed treatment assignment for all other treatments other than $m$ and $n$. In our simulation and case studies, upto five treatments options ($H=5$) are possible.

\subsubsection{Assumptions}
In applying Rubin's causal framework to estimate counterfactuals and label the treatment effect as causal, several critical assumptions must be made:
\begin{enumerate}
    \item Conditional exchangeability: This means the likelihood of receiving the treatment is based solely on observed covariates. Essentially, it assumes that any unmeasured factors influencing the outcome are equally distributed between the treated and untreated groups, provided that measured confounders are accounted for.
    \item Positivity: For every set of covariate values, there must be a nonzero chance of receiving each treatment condition (both treated and untreated). It's important that the outcome for an individual is independent of the treatment status of others.
    \item Consistency: The treatment should be clearly defined. This clarity is necessary to ensure that the level of exposure doesn't vary across subjects. Under this assumption, if the treatment is well-defined and consistent, we can infer that a subject's observed outcome would match their counterfactual outcome for their given exposure history.
\end{enumerate}

\subsection{Estimation methods}
G-formula, propensity score matching and targeted maximum likelihood estimation (TMLE) are popular statistical methods used in epidemiology and other fields for causal inference and adjusting for confounding in observational studies. Here is a brief introduction to each method and their extension to joint treatments and likert scale outcomes. 

\subsubsection{G-formula and modeling of likert scale outcomes}

\noindent The g-formula is an analytical tool for estimating standardized outcome distributions by using specific outcome distribution estimates based on covariates, which include both exposures and confounders \cite{hernan2010causal}. This formula can be applied to calculate common measures of association, such as the conditional risk differences (as in Equation \ref{equation1}). In our study, we evaluate the overall likert outcomes observed in our cohort and contrast it with the anticipated likert outcome in the same cohort if a new treatment had been administered.

Equation \ref{equation1} is a standard application of the g-formula, where we use regressions to estimate the potential likert outcomes under treatment and control conditions (i.e., $E(Y|W^{\eta}=1, X=x)$ and $E(Y|W^{\eta}=0, X=x)$), respectively. To construct the regression, we first standardize the individual likert outcomes which ranges from $1$ (Strongly Disagree) to $7$ (Strongly Agree) to a scale from 0 to 1 by performing the min-max transformation:
\begin{equation*}
    y_i^* = \frac{y_i-min(Y)}{max(Y)}, \text{ for i in 1,2, \ldots, n,}
\end{equation*}
\noindent where $min(Y)$ and max(Y) are the sample minimum and maximum outcomes. 

Hence, this can be seen as an extension of logistic regression \cite{rennie2005loss}, where typically only two categories, such as 'positive' and 'negative', are considered. However, our scenario includes more than two ordered levels. Similar to binary regression, we develop a real-valued predictor function $f(x)$ (for instance, a linear function in the case of linear binary regression), aiming to minimize a certain loss function, denoted as \( \text{loss}(f(x);y) \), against the target labels. One popular loss function is the logistic loss: 
\begin{equation}
\log(1 + \exp{f(x_i) y_i^*}),
\label{lossfunction}
\end{equation}
\noindent which measures the distance from the classification margin. In this context, we get the estimated potential outcome  $\hat{y_i}$ by converting the estimated outcome on probabilistic scale  $\hat{y_i}^*$  by applying the following formula: 

\begin{equation*}
\hat{y_i} = \hat{y_i}^* \cdot max(Y)+min(Y).
\end{equation*}

G-computation relies heavily on the correct specification of the outcome model. If the model does not accurately capture the relationship between the covariates, treatment, and outcome, the estimates produced may be biased. Selection bias can occur if certain key variables that influence both the selection into the study population and the outcome are omitted or incorrectly modeled.

\subsubsection{Propensity Score Matching}
Propensity score matching is a method to reduce selection bias in the estimation of causal effects. It involves matching treated and control units with similar propensity scores. It is intuitive and visually demonstrable, making it easier to communicate to non-statisticians. This method helps to create a balanced dataset by matching treated and untreated subjects with similar characteristics.

The propensity score is the probability of treatment assignment conditional on observed covariates, typically estimated using logistic regression:

\begin{equation}
    e(X,\eta) = P(W^{\eta}= 1 | X), 
\end{equation}

\noindent where $e(X,\eta)$ is the propensity score. 

After estimating the propensity scores, subjects are matched based on the euclidean distance of those scores to form comparable groups. One limitation of propensity score matching is that it can only balance observed covariates, leaving the possibility of bias due to unobserved confounders. Also, it can lead to a loss of data, as unmatched subjects are typically discarded.

\subsubsection{Targeted Maximum Likelihood Estimation (TMLE)}

TMLE is a semi-parametric estimation method that combines ideas from g-formula and propensity score adjustment \cite{van2006targeted}. It is a doubly robust estimation method because it has a unique property where the estimator can still be consistent and asymptotically normal if either the outcome or the propensity model is correctly specified.

TMLE involves iteratively updating an initial estimate of a statistical model (such as g-formula) to improve the estimation of a causal effect or a parameter of interest. The updating step is often represented as:
\begin{equation*}
Q_n^*(X) = Q_0(X) + \frac{I(W^{\eta}=1) - e(X,\eta)}{e(X,\eta)(1 - e(X,\eta))} \times (Y - Q_0(X))
\end{equation*}
\noindent where \( I(W^{\eta}=1) \) is an indicator function that equals 1 if the individual received the treatment and 0 otherwise. We update the initial outcome estimate $Q_0(X)$ to get a targeted outcome estimate $Q_n^*(X)$. This is done using a clever mechanism called the \textit{clever covariate}, denoted as $H(X)=1/e(X,\eta)$, which is essentially the inverse of the propensity score for each individual. The targeting step adjusts the initial estimate by a factor that depends on the clever covariate and the discrepancy between the observed outcome and the initial estimate. The TMLE estimate of the CATE, denoted as \( \hat{\Psi} \), can be computed as:
\begin{equation*}
   \hat{\Psi} = \frac{1}{n} \sum_{i=1}^{n} \left( Q_n^*(X=x_i | W^{\eta}=1) - Q_n^*(X=x_i | W^{\eta}=0) \right). 
\end{equation*}

TMLE's strength lies in its ability to handle complex, high-dimensional data and reduce bias in the estimation of causal effects, particularly in observational studies. It achieves this by iteratively refining the estimate to focus specifically on the parameter of interest, hence the name "targeted".

However, it's important to note that double robustness does not imply that TMLE is immune to all forms of bias. For instance, if both the outcome and treatment models are misspecified, TMLE estimates can still be biased. Additionally, TMLE, like any statistical method, is still susceptible to biases from unmeasured confounding, measurement error, and other sources of bias that are not related to the specific models used.

\subsubsection{CatBoost algorithm}
In our study, we employed the CatBoost algorithm \cite{prokhorenkova2018catboost} for fitting both the outcome model and the propensity models. CatBoost stands out in survey analysis due to its exceptional handling of categorical data, a common feature in surveys, without needing extensive preprocessing. It's highly robust against overfitting, crucial for complex datasets, and efficiently manages large datasets, making it ideal for extensive survey data. Additionally, its advanced treatment of missing values is particularly beneficial in survey contexts where such issues are prevalent. CatBoost also offers high accuracy in predictive modeling with features like ordered boosting, and despite its complexity, it provides interpretability through feature importance scores. This combination of efficiency, accuracy, and ease of integration with existing data analysis pipelines makes CatBoost a powerful and versatile tool in both survey analysis and regression tasks.

\subsection{Estimation performance using simulated dataset} 
To benchmark the algorithms, we simulate a dataset containing 5 covariates and 5 treatments, with the outcome variable being a Likert scale ranging from 1 to 7. The data simulation process is designed to explore the impact of these treatments and covariates on the outcome variable, providing insights into potential causal relationships.

The dataset consists of 5000 samples, each with 5 covariates \((X_1, X_2, X_3, X_4, X_5)\) generated from a normal distribution \(\mathcal{N}(1, 1)\). The treatment effects are uniformly set to 1 for all five treatments, indicating an equal impact of each treatment on the outcome variable. The treatment assignments are influenced by both the covariates and a confounding parameter, with a baseline probability adjusted based on the sum of the first three covariates and a confounding ratio.

The covariates for each sample are generated as follows:

\[
X_i \sim \mathcal{N}(1, 1), \quad i \in \{1, 2, 3, 4, 5\}
\]

where \(X_i\) represents the \(i\)-th covariate for each sample.

The probability of receiving each treatment is calculated using a combination of the covariates and a confounding factor (\(\text{confound}\)):

\[
\text{P}(T_j = 1) = \alpha \times 10^{-5} X + \beta \times \frac{\text{expit}\left(\sum_{k=1}^{3}X_k\right)}{2} + (1 - \beta) \times 0.5
\]
where:
\begin{itemize}
    \item $j \in \{1, 2, 3, 4, 5\}$, with only the first three covariates contributing to the unbalanced confounding,
    \item \(\alpha\) represents a small coefficient (e.g., \(10^{-5}\)) applied to the covariates,
    \item \(\beta\) corresponds to the confound ratio, it has been set to $0.15, 0.5$ and $0.85$ to indicate scenarios with low, medium and high level of confounding.
    \item \(\text{expit}(x) = \frac{1}{1 + e^{-x}}\) is the logistic function, and
    \item \(T_j\) is the treatment assignment for the \(j\)-th treatment.
\end{itemize}

The response variable (\(Y\)) is generated considering the baseline response, the effect of treatments, the impact of covariates, and random noise:

\[
Y = \text{clip}\left(1 + \sum_{j=1}^{5} T_j + \sum_{i=1}^{5} 0.1 X_i + \epsilon, 1, 7\right)
\]

where:
\begin{itemize}
    \item \(T_j\) denotes the effect of the \(j\)-th treatment,
    \item \(X_i\) represents the \(i\)-th covariate,
    \item \(\epsilon \sim \mathcal{N}(0, 0.1)\) is the noise term, and
    \item the clip function ensures $Y$ remains within the Likert scale range of 1 to 7.
\end{itemize}

We generate 50 sets of simulation datasets under each confounding scenario and compute the estimated conditional average treatment effect $\hat\Psi$ with the true treatment effect $\Psi$, which can be calculated using the counterfactual  $Y$ as explained in previous sections, the benchmark is computed as percentage error: $(\hat\Psi-\Psi)/\Psi$.

\subsubsection{Simulation Result}

\begin{figure*}[!h]
\centering 
\includegraphics[width = \linewidth]{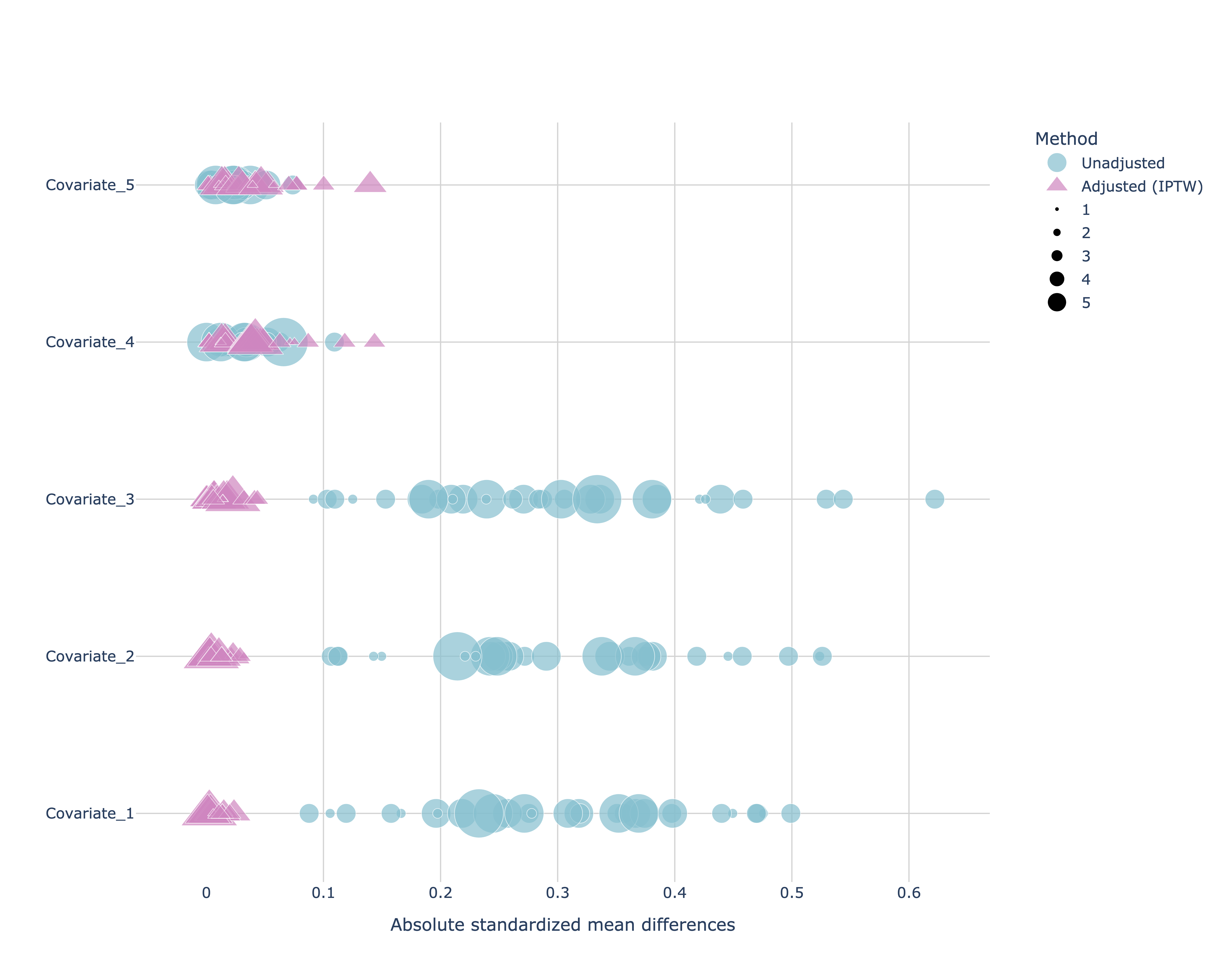}
\caption{Balance diagnostics showing absolute standardized mean differences for five covariates before and after adjustment. Each row represents a covariate with unadjusted differences displayed as blue circles and adjusted differences using inverse probability of treatment weighting (IPTW) shown as purple triangles. The degree of overlap indicates the effectiveness of the adjustment method in reducing bias across the covariates. The horizontal axis quantifies the magnitude of imbalance, with values closer to zero suggesting better balance between treatment groups.
}
\label{simulation2}
\end{figure*}

Figure \ref{simulation2}  illustrates a scenario of unbalanced confounding across five covariates from one random set of our simulations under medium confounding. In the unadjusted state, represented by blue circles, the absolute standardized mean differences for covariates 1, 2 and 3 indicate a substantial imbalance, suggesting that the treatment groups are systematically different regarding these covariates.

The application of inverse probability of treatment weighting (IPTW), depicted by purple triangles, aims to adjust for these differences. The effectiveness of this method can be evaluated by the proximity of the adjusted values to the baseline, indicating a mean difference of zero. IPTW adjustment reduces the imbalance in covariates with systematic confounding patterns, while keep others covariates (4 and 5) unadjusted, as evidenced by adjusted values that are still distant from the centerline.

\begin{figure*}[!h]
\centering 
\includegraphics[width = \linewidth]{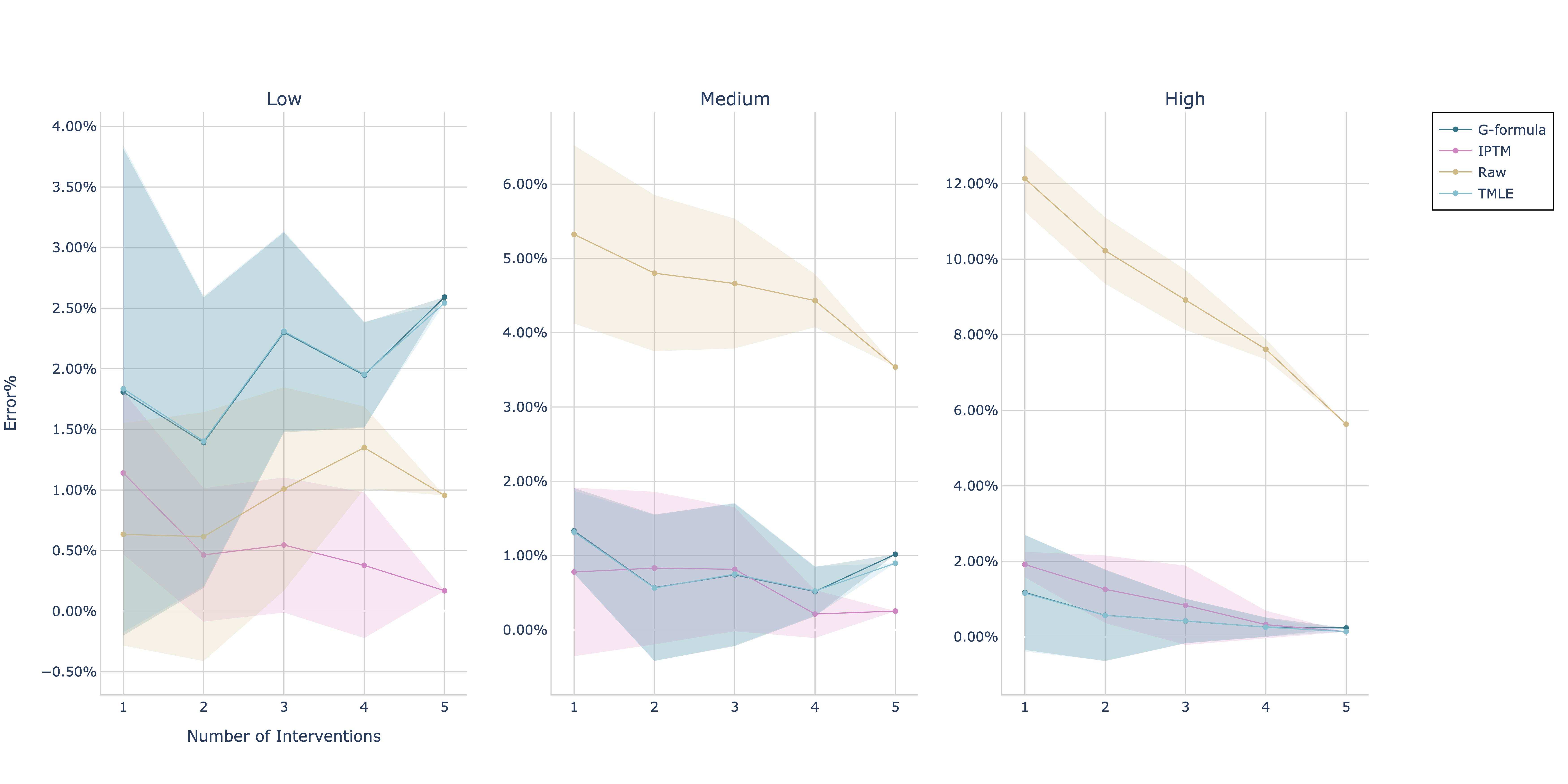}
\caption{Comparative analysis of conditional average treatment effect estimation methods across varying intervention scenarios. The three panels represent low, medium, and high levels of confounding, respectively. Within each panel, the estimation percentage error is plotted against the number of interventions, ranging from one to five. Estimation methods include the G-formula, IPTW, raw differences, and TMLE, each depicted by distinct colors and shapes. The shaded areas indicate the variability within the estimates, showcasing the robustness and sensitivity of each method under different confounding intensities.
}
\label{simulation}
\end{figure*}
To determine the most suitable approach for estimating the conditional treatment effect, we conducted a comparison of three popular methods in causal inference: G-formula \cite{hernan2010causal},  Propensity score matching \cite{austin2017performance} using simulated virtual experience data. 

Figure \ref{simulation} shows that in contexts of minimal confounding between the intervention and outcome, the inverse probability of treatment weighting (IPTW) method often outperforms the G-formula and targeted maximum likelihood estimation (TMLE) techniques. However, as confounding—or selection bias—grows more pronounced, the G-formula and TMLE demonstrate increased efficacy in treatment effect estimation. IPTW, while effective under certain conditions, shows limitations in both performance and its capacity to predict effects on new data due to reliance on matching within the observed dataset. Consequently, our analysis favors TMLE for its dual robustness, granting it an advantage over the G-formula in terms of accuracy in effect estimation.

\section{Results}\label{sec2}

\subsection{Imbalance in Confounders in Urban Design Studies}

\noindent Our investigation employed a virtual reality (VR) setup to elucidate the variance in walkability perceptions among distinct demographic groups. This analysis was predicated on the application of five targeted design environment interventions (as detailed in Table \ref{table_1}).

\begin{table}[!h]
\centering
\caption{Attributes and levels of the attributes.}
\begin{tabular}{ll}
\textbf{Intervention} & \textbf{Levels}                           \\
\hline
Land use mix (LM)         & (1) Residential land-use                  \\
                      & (0) Mixed with commercial area            \\
Block connectivity (BC)   & (1) High connectivity                     \\
                      & (0) Low connectivity                      \\
Road size    (RS)         & (1) Two lanes with narrow pedestrian zone \\
                      & (0) One lane with wide pedestrian zone    \\
Open space    (OS)        & (1) Has open space in the block           \\
                      & (0) Does not have open space in the block \\
Green/Trees        (GT)         & (1) Has trees in the block                \\
                      & (0) Does not have trees in the block      \\
\hline
\end{tabular}
\label{table_1}
\end{table}

\begin{figure*}[!htbp]
\centering 
\includegraphics[width = \linewidth]{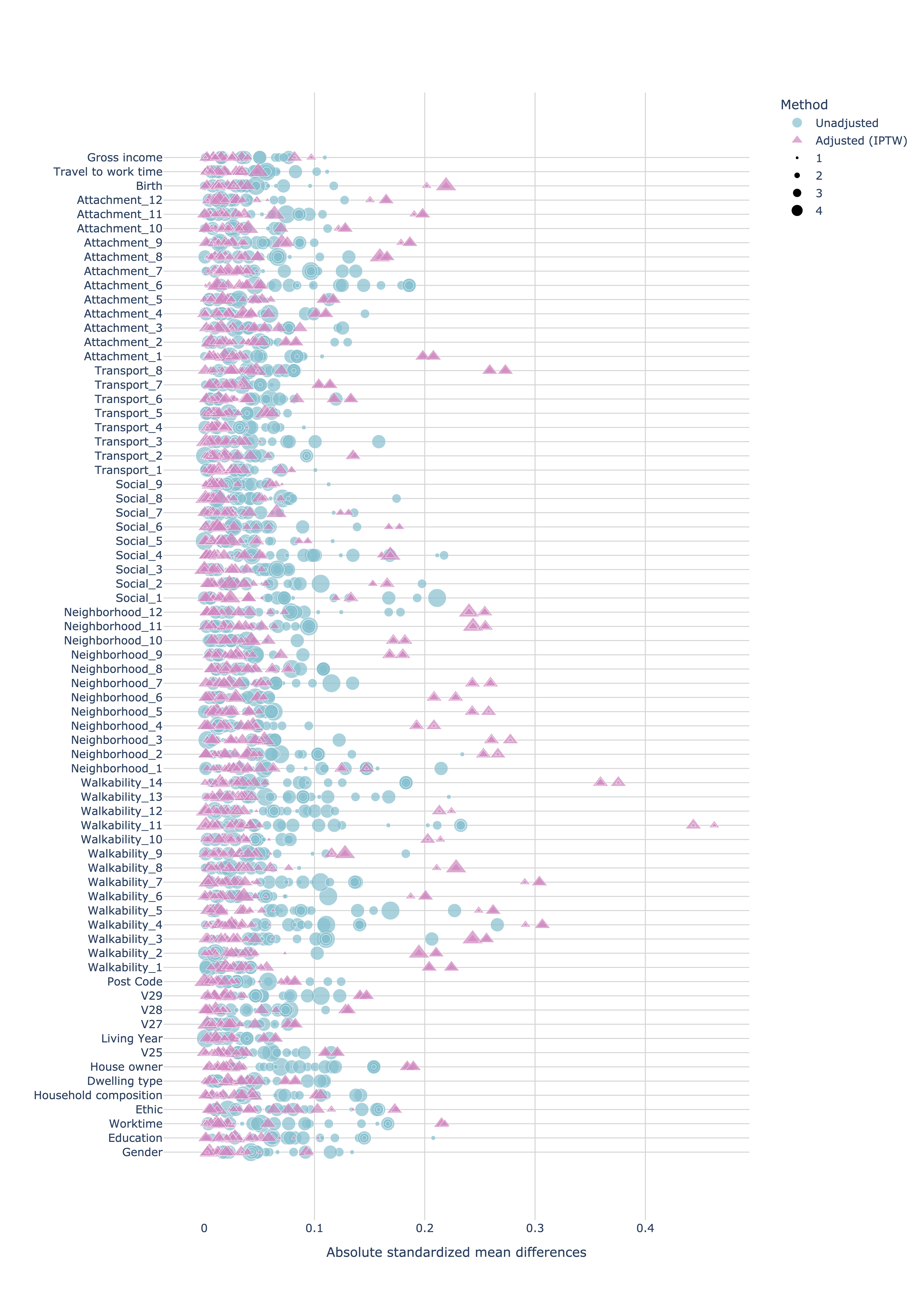}
\caption{Covariate balance illustration comparing pre- and post-inverse probability of treatment weighting (IPTW) adjustments. Each point represents the absolute standardized mean difference for a specific covariate in a dataset, with covariates listed on the y-axis. Covariates are organized into categories such as 'Gender', 'Ethnicity', 'Household Composition', and 'Walkability', among others. Two methods are depicted: unadjusted (circles) and adjusted (triangles) through IPTW, indicated by color and shape. The plot demonstrates the effectiveness of IPTW adjustment in reducing imbalance across the majority of covariates, with most post-adjustment differences clustering closer to zero, indicating improved balance. Marker size correlates with the number of interventions, providing insight into the relative weight each covariate holds within the adjusted model.}
\label{asmd}
\end{figure*}

The experimental framework facilitated the derivation of 32 ($\sum_{i=1}^{5}{C^5_n}$) possible combinations of conditional average treatment effects (CATE, refer to the Methodology section for a comprehensive explanation). Illustrated in Figure \ref{asmd}, the absolute standardized mean differences (ASMD) post-propensity score matching reveal a notable contraction in the discrepancies between covariates across treatment and control groups. This phenomenon indicates a moderate degree of confounding present within the VR dataset under examination.

The presence of unbalanced confounders between treatment and control groups introduces potential bias in the estimated CATE from our interventions, which may significantly affect the validity and reliability of causal inferences drawn from the study. 

In the context of our study, where diverse demographic groups' perceptions of walkability are evaluated through design interventions, such imbalances could skew the interpretation of how environmental changes impact pedestrian behavior. Specifically, if certain demographic characteristics or pre-existing conditions are disproportionately represented in either group, the observed effects may not accurately reflect the interventions' true impact but rather the underlying differences between the groups. This could lead to erroneous conclusions about the effectiveness of design interventions in improving urban walkability. To mitigate this issue, our study utilizes the Targeted Maximum Likelihood Estimation (TMLE) algorithm, an analytical approach designed for precision in the face of measured confounding variables.

\subsection{Conditional Average Treatment Effect}

The results presented in Table \ref{main_result} offer a nuanced understanding of how varied urban design interventions influence perceived walkability among different demographic segments.

With the application of a single intervention, Land use mix (LM) showed a modest but positive effect (CATE: 0.92\%, SE: 0.602\%), indicating that even isolated alterations to land use configurations can significantly enhance walkability perceptions. Block connectivity (BC) further exhibited a beneficial impact (CATE: 1.11\%, SE: 0.547\%), reinforcing the value of interconnected block designs in urban settings. In contrast, adjustments solely to Road size (RS) led to a marginal decrement in walkability (CATE: -1.21\%, SE: 0.489\%), suggesting that without additional supportive measures, enlarging road dimensions may adversely affect walkability.

Scenarios featuring Open space (OS) as the sole intervention demonstrated a marked positive effect (CATE: 3.83\%, SE: 0.716\%), underscoring the critical role open areas play in promoting walkability. Similarly, the exclusive addition of Green/Trees (GT) correlated with a favorable outcome (CATE: 2.03\%, SE: 0.525\%), highlighting the integral contribution of green elements to the urban environments.

Upon the integration of two or more interventions, combinations involving OS without LM yielded the highest CATE, approximately 4\%, manifesting the pronounced impact of open spaces when combined with other urban design elements, except LM, whereupon the CATE diminished to 3.486\%. Intriguingly, scenarios incorporating two or more interventions alongside LM experienced a decrease in CATE, illustrating a potential dilution or negative interaction effect when LM is part of multiple intervention strategies.

These findings suggest a complex interplay between various urban design interventions and their collective impact on walkability. The substantial positive effects of OS and GT, even as standalone interventions, reinforce the importance of integrating natural and open spaces within urban landscapes to enhance pedestrian experiences. 

Conversely, the observed decrease in CATE with certain combinations implies the need for careful consideration in selecting and implementing intervention mixtures to avoid counterproductive outcomes.

\begin{landscape}
\begin{table}[]
\resizebox{\linewidth}{!}{%
\begin{tabular}{@{}ccccccc|c@{}}
\toprule
\textbf{\#Interventions} & \textbf{Scenario} & \textbf{Land use mix (LM)} & \textbf{Block connectivity (BC)} & \textbf{Road size (RS)}   & \textbf{Open space (OS)}  & \textbf{Green/Trees (GT)} & \textbf{CATE}                 \\ \midrule
                         & 1                 & \cellcolor[HTML]{D0CECE}1  & 0                                & 0                         & 0                         & 0                         & $0.92\%^*$ \\
                         & 2                 & 0                          & \cellcolor[HTML]{D0CECE}1        & 0                         & 0                         & 0                         & $1.11\%^*$ \\
                         & 3                 & 0                          & 0                                & \cellcolor[HTML]{D0CECE}1 & 0                         & 0                         & $-1.21\%$                       \\
                         & 4                 & 0                          & 0                                & 0                         & \cellcolor[HTML]{D0CECE}1 & 0                         & $3.83\%^*$ \\
                          
\multirow{-5}{*}{1}      & 5                 & 0                          & 0                                & 0                         & 0                         & \cellcolor[HTML]{D0CECE}1 & $2.03\%^*$ \\
\hline
                         & 6                 & \cellcolor[HTML]{D0CECE}1  & \cellcolor[HTML]{D0CECE}1        & 0                         & 0                         & 0                         & $-0.10\%$                       \\
                         & 7                 & \cellcolor[HTML]{D0CECE}1  & 0                                & \cellcolor[HTML]{D0CECE}1 & 0                         & 0                         & $-2.47\%$                      \\
                         & 8                 & \cellcolor[HTML]{D0CECE}1  & 0                                & 0                         & \cellcolor[HTML]{D0CECE}1 & 0                         & $1.94\%^*$ \\
                         & 9                 & \cellcolor[HTML]{D0CECE}1  & 0                                & 0                         & 0                         & \cellcolor[HTML]{D0CECE}1 & $1.16\%^*$ \\
                         & 10                & 0                          & \cellcolor[HTML]{D0CECE}1        & \cellcolor[HTML]{D0CECE}1 & 0                         & 0                         & $-1.84\%$                       \\
                         & 11                & 0                          & \cellcolor[HTML]{D0CECE}1        & 0                         & \cellcolor[HTML]{D0CECE}1 & 0                         & $3.63\%^*$ \\
                         & 12                & 0                          & \cellcolor[HTML]{D0CECE}1        & 0                         & 0                         & \cellcolor[HTML]{D0CECE}1 & $1.74\%^*$ \\
                         & 13                & 0                          & 0                                & \cellcolor[HTML]{D0CECE}1 & \cellcolor[HTML]{D0CECE}1 & 0                         & $3.00\%^*$ \\
                         & 14                & 0                          & 0                                & \cellcolor[HTML]{D0CECE}1 & 0                         & \cellcolor[HTML]{D0CECE}1 & $2.03\%^*$ \\
\multirow{-10}{*}{2}     & 15                & 0                          & 0                                & 0                         & \cellcolor[HTML]{D0CECE}1 & \cellcolor[HTML]{D0CECE}1 & $4.75\%^*$ \\
\hline
                         & 16                & \cellcolor[HTML]{D0CECE}1  & \cellcolor[HTML]{D0CECE}1        & \cellcolor[HTML]{D0CECE}1 & 0                         & 0                         & $-1.79\%$                       \\
                         & 17                & \cellcolor[HTML]{D0CECE}1  & \cellcolor[HTML]{D0CECE}1        & 0                         & \cellcolor[HTML]{D0CECE}1 & 0                         & $2.47\%^*$ \\
                         & 18                & \cellcolor[HTML]{D0CECE}1  & \cellcolor[HTML]{D0CECE}1        & 0                         & 0                         & \cellcolor[HTML]{D0CECE}1 & $1.21\%^*$ \\
                         & 19                & \cellcolor[HTML]{D0CECE}1  & 0                                & \cellcolor[HTML]{D0CECE}1 & \cellcolor[HTML]{D0CECE}1 & 0                         & $1.74\%^*$ \\
                         & 20                & \cellcolor[HTML]{D0CECE}1  & 0                                & \cellcolor[HTML]{D0CECE}1 & 0                         & \cellcolor[HTML]{D0CECE}1 & $1.02\%^*$ \\
                         & 21                & \cellcolor[HTML]{D0CECE}1  & 0                                & 0                         & \cellcolor[HTML]{D0CECE}1 & \cellcolor[HTML]{D0CECE}1 & $2.52\%^*$ \\
                         & 22                & 0                          & \cellcolor[HTML]{D0CECE}1        & \cellcolor[HTML]{D0CECE}1 & \cellcolor[HTML]{D0CECE}1 & 0                         & $3.20\%^*$ \\
                         & 23                & 0                          & \cellcolor[HTML]{D0CECE}1        & \cellcolor[HTML]{D0CECE}1 & 0                         & \cellcolor[HTML]{D0CECE}1 & $1.07\%^*$ \\
                         & 24                & 0                          & \cellcolor[HTML]{D0CECE}1        & 0                         & \cellcolor[HTML]{D0CECE}1 & \cellcolor[HTML]{D0CECE}1 & $3.97\%^*$ \\
\multirow{-10}{*}{3}     & 25                & 0                          & 0                                & \cellcolor[HTML]{D0CECE}1 & \cellcolor[HTML]{D0CECE}1 & \cellcolor[HTML]{D0CECE}1 & $4.65\%^*$ \\
\hline
                         & 26                & \cellcolor[HTML]{D0CECE}1  & \cellcolor[HTML]{D0CECE}1        & \cellcolor[HTML]{D0CECE}1 & \cellcolor[HTML]{D0CECE}1 & 0                         & $1.40\%^*$ \\
                         & 27                & \cellcolor[HTML]{D0CECE}1  & \cellcolor[HTML]{D0CECE}1        & \cellcolor[HTML]{D0CECE}1 & 0                         & \cellcolor[HTML]{D0CECE}1 & $0.44\%$                        \\
                         & 28                & \cellcolor[HTML]{D0CECE}1  & \cellcolor[HTML]{D0CECE}1        & 0                         & \cellcolor[HTML]{D0CECE}1 & \cellcolor[HTML]{D0CECE}1 & $3.58\%^*$ \\
                         & 29                & \cellcolor[HTML]{D0CECE}1  & 0                                & \cellcolor[HTML]{D0CECE}1 & \cellcolor[HTML]{D0CECE}1 & \cellcolor[HTML]{D0CECE}1 & $3.39\%^*$ \\
                          
\multirow{-5}{*}{4}      & 30                & 0                          & \cellcolor[HTML]{D0CECE}1        & \cellcolor[HTML]{D0CECE}1 & \cellcolor[HTML]{D0CECE}1 & \cellcolor[HTML]{D0CECE}1 & $4.65\%^*$ \\
\hline
5                        & 31                & \cellcolor[HTML]{D0CECE}1  & \cellcolor[HTML]{D0CECE}1        & \cellcolor[HTML]{D0CECE}1 & \cellcolor[HTML]{D0CECE}1 & \cellcolor[HTML]{D0CECE}1 & $3.49\%^*$ \\ \bottomrule
\end{tabular}
}
\caption{The estimated conditional average treatment effect of various intervention combinations on the Likert-scale outcome.; *: the effect is significant at $95\%$ level}
\label{main_result}
\end{table}
\end{landscape}

\subsection{Effect Sensitivity}

\begin{figure*}[!htbp]
\centering 
\includegraphics[width = \linewidth]{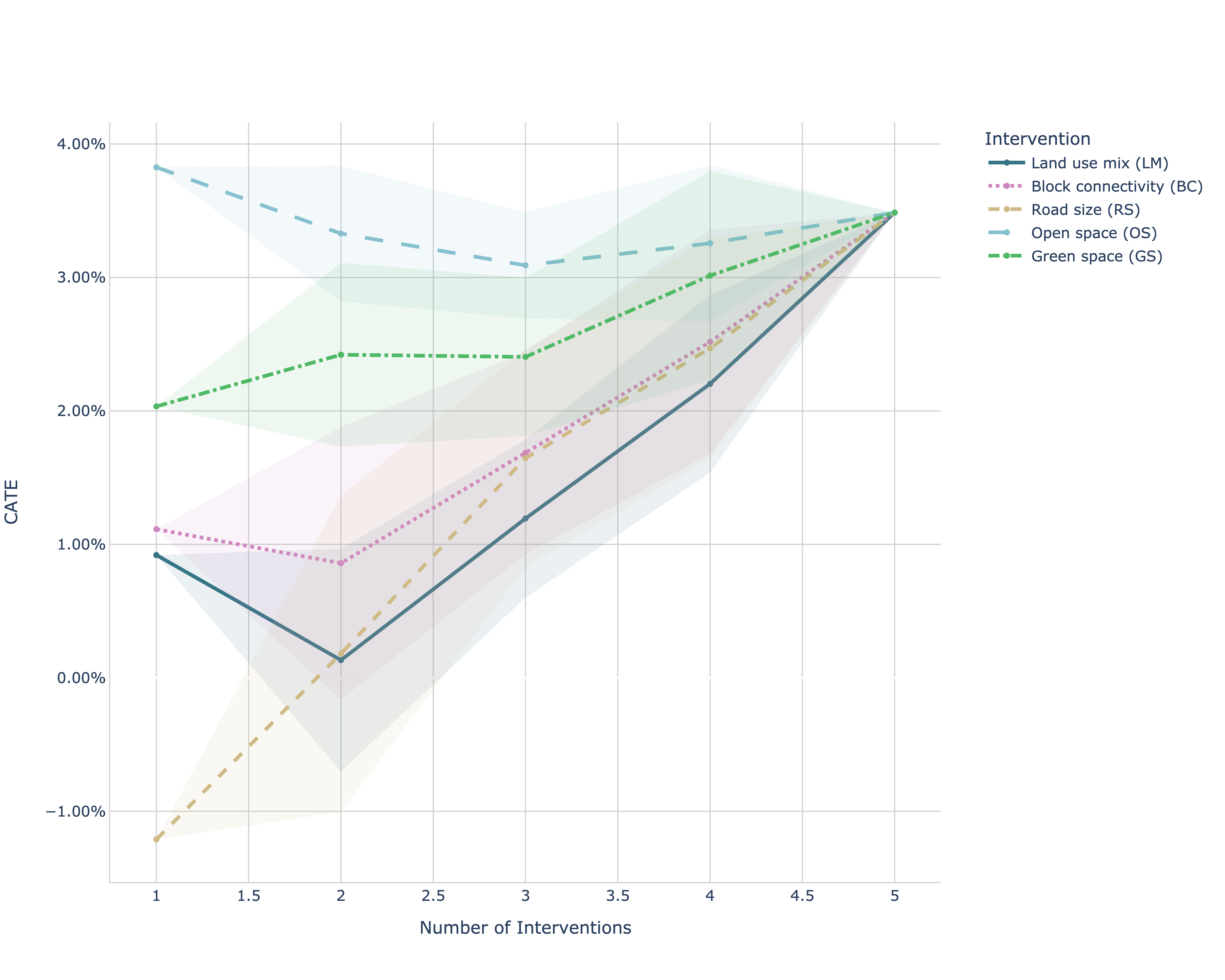}
\caption{The Conditional Average Treatment Effect (CATE) as a function of the number of interventions for five interventions: land use mix, high block connectivity, two lanes, open space, and the presence of trees. Each trace corresponds to the CATE for selected interventions applied to each urban feature, averaged over all other confounders and interventions, plotted against the number of coexist-interventions. The shaded areas around the traces indicate the 95\% confidence intervals, which delineate the precision of the estimated treatment effect at each level of intervention. This visualization enables the assessment of how different urban planning features influence the CATE, providing insights into their potential impact within the urban context.
}
\label{benchmark1}
\end{figure*}

In Figure \ref{benchmark1}, we present an analysis of the impact of urban design interventions, contingent on the extent of concurrent interventions. The number of interventions—ranging from one, indicating the exclusive presence of the intervention under consideration, to five, denoting the hypothetical application of all interventions—serves as a proxy for the complexity of urban design strategies. The Conditional Average Treatment Effect (CATE) is computed for select interventions, adjusting for potential confounders and the influence of other concurrent interventions based on Table \ref{main_result}. It reveals a spectrum of responsiveness across interventions: certain measures are more heavily influenced by the number of concurrent interventions than others. This variability underscores a complex interplay among urban design measures and highlights the importance of a calibrated approach in policy formulation and the execution of urban development agendas.

First, the analysis reveals distinct patterns of influence as the number of interventions increases. LM shows an incremental positive trend in average CATE values with additional interventions, starting from a negligible effect at a single intervention (CATE = 0.92\%) to a substantially larger effect with five interventions (CATE = 3.49\%). This pattern suggests that interventions promoting mixed land use could have compounding beneficial impacts.

Notably, OS initially indicates an effect (CATE = 3.83\%) for a single intervention, yet keeps stable above $3\%$ as the number of interventions rises. It consistently exhibits the highest CATE values across all levels of intervention, implying that creating or preserving open spaces could be the most effective singular urban design intervention in our case study.

RS, BC and GT parallels LM with increasing positive CATE values, while GT demonstrates a unique profile with the CATE starting at a higher baseline (CATE = 2.03\% for a single intervention) and also enhancing as the number of interventions grows. The precision of these effects is demonstrated by the 95\% confidence intervals, which are notably tight for OS, suggesting a high level of confidence in its positive impact.

Overall, the evidence advocates for a strategic approach to urban design, where specific combinations and numbers of interventions can be tailored to maximize the beneficial effects of urban features on the targeted outcomes.

\section{Discussion}\label{sec12}

The present study combines a VR-based conjoint experiment with heterogeneous treatment effect estimation to investigate the impact of urban design interventions on walkability. Recent studies have highlighted the value of VR as a tool in urban design and planning research, as it allows for the controlled manipulation of environmental variables and the effective assessment of user responses in a safe and cost-effective manner\cite{yin2015big, lu2021finding, jiang2023factors, qiao2023understanding, ferre2022adoption}. Our research extends these findings by utilizing a VR-based conjoint experiment to explore how demographic characteristics influence perceived walkability under various urban design scenarios. 

Employing Targeted Maximum Likelihood Estimation (TMLE) for Conditional Average Treatment Effect (CATE) estimation marks a significant methodological advancement in urban design and planning research, particularly in addressing the challenge of imbalanced confounders\cite{borah2013highlighting, kang2016instrumental, liang2021analyzing}. This approach, celebrated for its double robustness, significantly mitigates bias and enhances efficiency, enabling nuanced analyses of urban design interventions on walkability across varied demographic groups\cite{van2011targeted, schuler2017targeted, chen2022predicting, guo2013pedestrian, li2015assessing}. Our integration of TMLE within a VR-based experimental framework underscores the potential to refine research validity and reliability in urban studies. The resulting CATE estimations provide insights into treatment effect heterogeneity, guiding the development of targeted urban design strategies that meet the diverse needs of different neighborhoods, thereby facilitating informed decision-making and resource optimization in urban design and planning\cite{chen2019discovering, chen2022predicting, liang2023revealing, beck2023developing, liu2019exploring}. This approach not only enhances our understanding of urban walkability design interventions' varied impacts but also supports the creation of more inclusive and effective urban walkable environments.

Our results presents a vivid depiction of the concurrent urban design interventions' impact on walkability, echoing the established narrative that pedestrian experiences are influenced by a complex array of urban elements\cite{ewing2010travel, sung2013evidence}. The observed enhancements from individual interventions like Land use mix (LM) and Block connectivity (BC) are in line with Ewing and Cervero's "5Ds" framework, highlighting critical aspects such as density and diversity for fostering walkability\cite{ewing2010travel}. Conversely, the study underscores the potential drawbacks of enlarging Road size (RS) on its own, supporting Guo and Loo's critique of road expansions harming pedestrian safety and comfort\cite{guo2013pedestrian}.

The beneficial effects of integrating Open space (OS) and Green/Trees (GT) interventions support the notion, underpinned by the Biophilia hypothesis, that natural elements within urban settings significantly enhance walkability by catering to humans' inherent affinity for nature\cite{wilson1986biophilia, koohsari2015public, lu2018effect,liang2021analyzing,wang2021influence, polson2017deep}. This enhancement is attributed to the role of OS and GT in improving safety perceptions, alleviating stress, and fostering social connections\cite{beck2023developing,qiao2023understanding,fancello2023micro}, highlighting their strategic value in urban design.

Moreover, this study illuminates the intricate dynamics between various urban design interventions, revealing that certain combinations, especially those including LM, may lead to diminished CATE values, indicating potential counterproductive effects or interactions. This complexity underscores the necessity for a contextually informed and strategic urban design approach, as advocated by Sallis et al. (2016) regarding physical activity and built environments\cite{sallis2016physical}. The continued exploration into the nuanced relationships among urban design features and their collective impact on walkability\cite{beck2023developing, qiao2023understanding, fancello2023micro, liang2023revealing, ferre2022adoption, liang2021analyzing, wang2021influence} calls for a refined understanding that can guide the creation of more targeted, effective urban development strategies. By integrating advanced statistical methodologies like TMLE and leveraging findings from recent research, urban planners and policymakers are better positioned to cultivate walkability across varied urban environments.

%Limitations and Future Directions

However, it is essential to acknowledge the potential limitations of our study. While the VR-based experimental setup allows for a controlled manipulation of environmental factors, it may not fully capture the dynamic nature of real-world urban environments and the long-term effects of interventions. Additionally, the TMLE algorithm, although robust, relies on the correct specification of the treatment or outcome models\cite{van2011targeted}. Misspecification of both models may lead to biased estimates, highlighting the importance of careful model selection and sensitivity analyses. Future research should focus on validating the findings of our study in real-world settings, employing longitudinal designs and field experiments to assess the long-term impact of urban design interventions on perceived walkability and walking behaviors. Incorporating advanced data collection methods, such as long-tracking measurements and ecological momentary assessments\cite{chaix2012interactive, shiffman2008ecological, fancello2023micro}, can provide a more comprehensive understanding of the dynamic interactions between individuals and the built environments.

The integration of TMLE with cutting-edge machine learning algorithms and sophisticated spatial analysis methodologies presents a promising avenue for future research to improve the precision and dependability of results\cite{athey2016recursive, kang2016instrumental}. The field of machine learning has recently made significant strides, revealing intricate patterns and interconnections within urban datasets\cite{polson2017deep, qiao2023understanding}. These breakthroughs afford scholars a more comprehensive insight into the intricate nexus of urban planning, walkability, and public health outcomes. Employing spatial analytical techniques, such as geographically weighted regression (GWR) and spatial autocorrelation analysis, is critical for recognizing the spatial dependencies and variations that are characteristic of urban data.

In conclusion, our study leverages the potential of VR technology and advanced statistical methods to explore urban design interventions and their effects on walkability. Through controlled virtual experiments and sophisticated data analysis, we illuminate the complex interplay between urban design elements and their impact on different demographic groups. Our findings underscore the importance of integrating natural spaces, enhancing connectivity, and carefully considering the scale of roadways to foster walkable urban environments. By doing so, our research contributes valuable insights into the optimization of urban spaces that cater to the diverse needs and preferences of city dwellers. Looking forward, it is clear that the intersection of technological innovation and methodological rigor will continue to refine our understanding the role of urban designer and planner in promoting public health and sustainability. As we advance, embracing the complexity of urban environments through comprehensive, data-driven approaches will be paramount in designing cities that are not only walkable but also inclusive, resilient, and conducive to well-being.

\newpage
\backmatter

\bmhead{Data availability}
The data that support the findings of this study are available from the corresponding author, [Bojing Liao], upon reasonable request.

\bmhead{Declarations}
No conflict of interest declared by the authors. All authors agreed to the submission of this version of the manuscript and are responsible for its content.

\bmhead{Funding}
The present study is supported by the Fundamental Research Funds for the Central Universities (with grant number 20720221045), and the Guangdong Basic and Applied Basic Research Foundation (with grant number 2023A1515110663).

\newpage
\begin{appendices}

\section{Conditional average treatments effect estimation for 2 or more interventions}
The interaction among interventions can be represented by the following CATEs: 

\begin{align*}
    \psi^{LM}(X|BC=1) &= E_{X=x}\bigg(E(Y|LM=1,BC=1,X=x) - E(Y|LM=0,BC=1,X=x)\bigg)\\
    \psi^{LM}(X|BC=0) &= E_{X=x}\bigg(E(Y|LM=1,BC=0,X=x) - E(Y|LM=0,BC=0,X=x)\bigg)\\
    \psi^{BC}(X|LM=1) &= E_{X=x}\bigg(E(Y|LM=1,BC=1,X=x) - E(Y|LM=1,BC=0,X=x)\bigg)\\
    \psi^{BC}(X|LM=0) &= E_{X=x}\bigg(E(Y|LM=0,BC=1,X=x) - E(Y|LM=0,BC=0,X=x)\bigg)
\end{align*}
Here, we notice that when there is no interaction between two interventions, we have 
\[
\psi^{LM}(X|BC=1) = \psi^{LM}(X|BC=0) = \psi^{LM}(X).
\]

\noindent Otherwise, we have: 
\begin{equation*}
 \psi(X)^{LM\&BC} = 1/2\big{(}\psi^{LM}(X|BC=1)+\psi^{LM}(X|BC=0)+ \psi^{BC}(X|LM=1)+\psi^{BC}(X|LM=0)\big{)}
\end{equation*}

\textbf{Proof:}

\begin{align*}
 &\psi^{LM}(X|BC=1)+\psi^{LM}(X|BC=0)+ \psi^{BC}(X|LM=1)+\psi^{BC}(X|LM=0) \\
= &E_{X=x}(\\
&E(Y|LM=1,BC=1,X=x) - E(Y|LM=0,BC=1,X=x)+\\
&E(Y|LM=1,BC=0,X=x) - E(Y|LM=0,BC=0,X=x)+\\
&E(Y|LM=1,BC=1,X=x) - E(Y|LM=1,BC=0,X=x)+\\
&E(Y|LM=0,BC=1,X=x) - E(Y|LM=0,BC=0,X=x))\\
=&E_{X=x}(\\
&E(Y|LM=1,BC=1,X=x)-E(Y|LM=0,BC=0,X=x)+\\
&E(Y|LM=1,BC=1,X=x)-E(Y|LM=0,BC=0,X=x))\\
&=2E_{X=x}(E(Y|LM=1,BC=1,X=x)-E(Y|LM=0,BC=0,X=x))\\
&=2\psi(X)^{LM\&BC}
\end{align*}

\qed

\end{appendices}

\bibliography{bibliography}
\end{document}